\documentclass[%
reprint,
superscriptaddress,
showpacs,preprintnumbers,
nofootinbib,
amsmath,amssymb,
aps,
prd,
floatfix,
]{revtex4-1}

\usepackage{graphicx}
\usepackage{dcolumn}
\usepackage{bm}
\usepackage{hyperref}
\usepackage[mathlines]{lineno}
\usepackage{float}
\usepackage{color}

\begin{document}


\title{Microlensing and dynamical constraints on primordial black hole dark matter with an extended mass function}

\author{Anne M. Green}
\email{anne.green@nottingham.ac.uk}
\affiliation{School of Physics and Astronomy, University of Nottingham, University Park, Nottingham, NG7 2RD, United Kingdom}

\date{\today}

\begin{abstract}
The recent discovery of gravitational waves from mergers of $\sim 10 \, M_{\odot}$  black hole binaries has stimulated interested in Primordial Black Hole dark matter in this mass range. Microlensing and dynamical constraints exclude all of the dark matter being in compact objects with a delta function mass function in the range $10^{-7} \lesssim M/ M_{\odot} \lesssim 10^{5}$. However it has been argued that all of the dark matter could be composed of compact objects in this range with an extended mass function. We explicitly recalculate the microlensing and dynamical constraints for compact objects with an extended mass function which replicates the PBH mass function produced by inflation models. We find that the microlensing and dynamical constraints
place conflicting constraints on the width of the mass function, and do not find a mass function which satisfies both constraints.
\end{abstract}

\maketitle

\section{Introduction}

Primoridal black holes (PBHs) can form in the early Universe via the collapse of large density perturbations~\cite{Carr:1974nx,Carr:1975qj} produced by a period of inflation~\cite{Carr:1993aq}. PBHs with mass $M \gtrsim 10^{15} \, {\rm g}$ will not have evaporated by the present day~\cite{Carr:1976zz}. Since PBHs form before nucleosynthesis they are non-baryonic and are therefore a cold dark matter candidate~\cite{Carr:2016drx}. 
There are various, mass dependent, constraints on the abundance of PBHs, from their lensing and dynamical effects, and also their effects on various astrophysical objects and processes. See Refs.~\cite{Carr:2009jm,Green:2014faa,Carr:2016drx} for compilations of these constraints. 

Ref.~\cite{Carr:2016drx} has highlighted three mass windows (at $10^{16}-10^{17} \, {\rm g}$, $10^{20} -10^{24} \, {\rm g}$ and $1-10^3 \, M_{\odot}$) 
where PBHs could potentially make up all of the dark matter. The observational constraints on the abundance of PBHs are usually calculated
assuming a delta function mass function, and exclude all of the dark matter being in PBHs of any single mass. Refs.~\cite{Clesse:2015wea,Carr:2016drx} have pointed out that an extended mass function, as is expected to be produced from the collapse of large inflationary density perturbations~\cite{Niemeyer:1997mt,Kuhnel:2015vtw}, might still be compatible with all of the observational constraints. The constraints calculated assuming a delta function mass function can not be directly applied to an extended mass function however. In this paper we explicitly recalculate the constraints for extended mass functions which mimic those
expected for PBHs produced by the collapse of inflationary density perturbations. We focus on the intermediate black hole mass range, $1-10^3 \, M_{\odot}$, since there has been much recent interest in PBHs of this mass~\cite{Bird:2016dcv,Clesse:2016vqa,Sasaki:2016jop} in light of the discovery of gravitational waves from $\sim 10 \, M_{\odot}$ BH binaries by LIGO~\cite{Abbott:2016blz}.

In Sec.~\ref{sec:constraints} we review the microlensing~\cite{Tisserand:2006zx} and dynamical~\cite{Brandt:2016aco} constraints on intermediate mass 
($1 \lesssim M/M_{\odot} \lesssim 10^{3}$ MAssive Compact Halo Objects. In Sec.~\ref{sec:emf} we recalculate these constraints for extended differential halo fractions (DHFs) which mimic the DHFs found for PBHs produced from the collapse of large inflationary density perturbations in Ref.~\cite{Carr:2016drx}. Finally we conclude with discussion in Sec.~\ref{sec:discussion}.

\section{Constraints}
\label{sec:constraints}

We will consider the same two constraints considered in Ref.~\cite{Carr:2016drx} for intermediate mass PBHs: microlensing~\cite{Tisserand:2006zx} and dynamical heating of star clusters/ultra-faint dwarf galaxies~\cite{Brandt:2016aco}. There are other constraints on compact objects of this mass. 
Disruption of wide binaries exclude halo fractions, $f=\rho_{\rm MACHO}/\rho_{\rm DM}$, greater than unity for $M \gtrsim 100 \, M_{\odot}$~\cite{Chaname:2003fn,Yoo:2003fr,Quinn:2009zg,Monroy-Rodriguez:2014ula}. However these constraints require assumptions about the initial distribution of the semi-major axes of the binaries, and also the smallest MACHO mass for which a delta function mass function is excluded depends on what sub-set of binaries is considered~\cite{Monroy-Rodriguez:2014ula}.
The X-rays emitted due to accretion of gas onto multi-Solar mass PBHs would produce measurable changes in the spectrum and anisotropies of the Cosmic Microwave Background radiation~\cite{Ricotti:2007au,Chen:2016pud}. However Ref.~\cite{Bird:2016dcv} argues that there are significant uncertainties associated with modelling the complex physical processes involved. The strong gravitational lensing of extragalactic fast radio bursts will place tight limits
on MACHOs in this mass range in the future~\cite{Munoz:2016tmg}.

\subsection{Microlensing}
\label{sec:micro}

Microlensing is the temporary amplification of a background star which occurs when a compact object passes close to the line of sight to the background star~\cite{Paczynski:1985jf}.
A microlensing event occurs when a compact object passes through the microlensing `tube', which has a radius of 
$u_{{\rm T}} R_{{\rm E}}$ where $u_{{\rm T}} \approx 1$ is the minimum impact
parameter for which the amplification of the background star is above
the required threshold and $R_{{\rm E}}$ is the Einstein radius:
\begin{equation}
R_{{\rm E}}(x)= 2 \left[ \frac{ G M x (1-x)L}{c^2 } \right]^{1/2} \,,
\end{equation}
where $L$ is the distance to the source, $M$ is the MACHO mass and $x$ is the distance of the MACHO from the observer, in units of $L$~\cite{Paczynski:1985jf}. 
The distance to the LMC is much greater than its line of sight depth, so all of the source stars can be assumed to be at the same distance ($\sim 50$ kpc) and the angular distribution of sources ignored. 
For a non--delta function mass function $\psi(M)$, defined so that the fraction, $f$, of the total mass of the halo in the form of MACHOs is
\begin{equation}
 f= \int_{0}^{\infty} \psi(M) \, {\rm d} M \,,
 \end{equation}
the differential event rate is~\cite{Griest:1990vu,DeRujula:1990wq,Alcock:1996yv}~\footnote{This expression assumes a a spherical halo with an isotropic
velocity distribution and ignores the transverse velocity of the microlensing tube, which has a small effect on the differential event rate~\cite{Griest:1990vu}.}:
\begin{equation}
\label{df}
\frac{{\rm d} \Gamma}{{\rm d} \hat{t}} =  \frac{32 L u_{{\rm T}} }
                 {{\hat{t}}^4
              {v_{{\rm c}}}^2}
                 \int_{0}^{\infty}  \left[ \frac{\psi(M)}{M}  
              \int^{x_{{\rm h}}}_{0} \rho(x) R^{4}_{{\rm E}}(x)
              e^{-Q(x)}  {\rm d} x \right] {\rm d} M \,, 
\end{equation}
where $\hat{t}$ is the time taken to cross the Einstein {\it diameter}, $x_{{\rm h}} \approx 1$ is the extent of the halo and 
$Q(x)= 4 R^{2}_{{\rm E}}(x) u_{{\rm T}}^2 / (\hat{t}^{2} v_{{\rm c}}^2)$, where $v_{\rm c}  =220 \, {\rm km \, s}^{-1}$ is the local circular speed. 

Microlensing analyses usually assume a standard halo, which consists of a cored isothermal sphere:
\begin{equation}
\rho(R) = \rho_{0} \frac{R_{{\rm c}}^2 + R_{0}^2}{R_{{\rm c}}^2 + R^2} \,,
\end{equation}
with local dark matter density $\rho_{0}= 0.0079 M_{\odot} {\rm pc}^{-3}$, core radius $R_{{\rm c}} \approx 5$ kpc  and
Solar radius $R_{0} \approx 8.5$ kpc. Eq.(\ref{df}) then becomes~\cite{Alcock:1996yv}
\begin{eqnarray}
\frac{{\rm d} \Gamma}{{\rm d} \hat{t}} &=& \frac{512 \rho_{0} 
          (R_{{\rm c}}^2 + R_{0}^2) L G^2 u_{{\rm T}} }
             {{\hat{t}}^4 {v_{{\rm c}}}^2 c^4}
            \nonumber \\ 
      \times   \int_{0}^{\infty} && \hspace{-0.5cm} \left[ \psi(M) M  
              \int^{x_{{\rm h}}}_{0} \frac{x^2 (1-x)^2}{A + B x + x^2}
            e^{-Q(x) } {\rm d} x \right] {\rm d} M \,, 
\end{eqnarray}
where $A=(R^2_{{\rm c}}+ R^2_{0})/L^2$, $B=-2(R_{0}/L) \cos{b}
\cos{l}$ and $b=-33^{\circ}$ and $l=280^{\circ}$ are the galactic
latitude and longitude, respectively, of the LMC.

\begin{figure}
\includegraphics[width=0.45\textwidth]{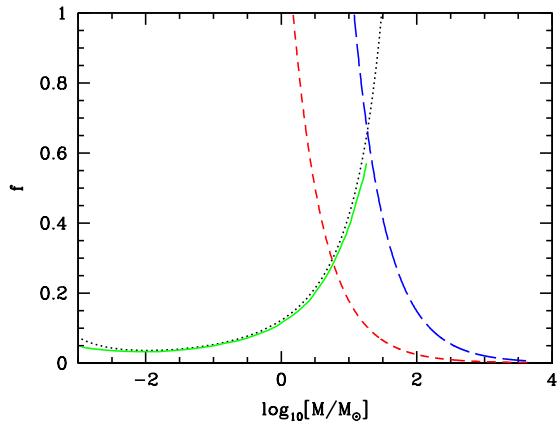}
\caption{\label{fig:EROS} The constraints on the halo fraction, $f$, of MACHOs as a function of mass, $M$, for a delta function mass function. The green solid line is a digitisation of the result from the EROS-2 microlensing survey (bottom panel of Fig.~15 of Ref.~\cite{Tisserand:2006zx})
 and the black dotted line is our implementation of their constraint, as described in the text. The published EROS-2 limit stops at $f \sim 0.6$ because they only plot the limit for halo fractions in the range $0<f<0.6$. The short red and long blue dashed lines are from the disruption of the star cluster in Eridanus II and the heating of ultra-faint dwarfs respectively~\cite{Brandt:2016aco}. See the text for details }
\end{figure}

The expected number of events, $N_{\rm exp}$, is given by
\begin{equation}
N_{{\rm exp}} = E \int_{0}^{\infty} \frac{{\rm d} \Gamma}{{\rm d} \hat{t}}
           \,  \epsilon(\hat{t}) \, {\rm d} \hat{t} \,,
\end{equation}
where $E$ is the exposure in star years and $\epsilon(\hat{t})$ is the detection efficiency i.e. the probability that a microlensing event with duration $\hat{t}$ is detected. For the EROS-2 survey
$E=3.77 \times 10^{7}$ star years. The detection efficiency, in terms of Einstein radius crossing time, is given in Fig.~11 of Ref.~\cite{Tisserand:2006zx} (and as stated in the figure caption is multiplied by a factor of 0.9 to take into account lensing by binary lenses). No events were observed and the EROS collaboration calculate constraints on the halo fraction, $f$, for a delta function mass function (i.e. with $\psi(M) =\delta(M)$) by finding the value of $f$ for which $N_{\rm exp} = 3.0$. Their results, from Fig.~15 of Ref.~\cite{Tisserand:2006zx}, are shown in Fig.~\ref{fig:EROS} along with our implementation of their constraints. Our constraints are in good agreement with theirs for $-3 < \log_{10} {(M/M_{\odot})} < 1$. Tighter constraints can be obtained for $\log_{10}(M/M_{\odot})< -3 $ by considering the EROS-1 and MACHO data~\cite{Alcock:1998fx}, however these data do not place tighter constraints on the multiple Solar mass PBHs that we are interested in. Marginally tighter constraints could be obtained for $\log_{10}(M/M_{\odot})>  -2 $ by also considering the EROS-2 SMC data. However the one observed event is consistent with expectations from self-lensing (i.e. lensing by objects in the SMC itself)~\cite{Graff:1998ix}.

\subsection{Dynamical constraints}
\label{sec:dyn}

Ref.~\cite{Brandt:2016aco} showed that MACHOs with $M \gtrsim 5 M_{\odot}$ dynamically heat the stars in star clusters or ultra-faint dwarf galaxies causing the half-light radius, $r_{\rm h}$, to evolve with time as
\begin{equation}
\label{rh}
 \frac{{\rm d} r_{\rm h}}{{\rm d} t} = \frac{ 4 \sqrt{2} \pi G f M}{\sigma} \ln{\Lambda} \left( \alpha \frac{M_{\star}}{\rho r_{\rm h}^2} + 2 \beta r_{\rm h} \right)^{-1} \,, 
\end{equation}
where $\sigma$ is the MACHO velocity dispersion, $M_{\star}$ is the total stellar mass, $\rho$ is the total dark matter density, $\alpha$ and $\beta$ are parameters that depend on the mass distribution and $\ln{\Lambda}$ is the Coulomb logarithm
\begin{equation}
\ln{\Lambda} \approx \ln{ \left( \frac{ r_{\rm h} \sigma^2}{G (M_{\odot} + M)} \right)} \,.
\end{equation}
We follow Ref.~\cite{Brandt:2016aco} and take $\alpha=1$ and $\beta=10$.

Ref.~\cite{Brandt:2016aco} found the constraints on the MACHO halo fraction, $f$, as a function of MACHO mass, $M$, from the disruption of the $M_{\star} = 3000 \, M_{\odot}$ star cluster~\cite{crnojevic} 
at the centre of Eridanus II~\cite{Koposov:2015cua,Bechtol:2015cbp} and also from the observed sizes of the compact ultra-faint dwarf galaxies~\cite{Simon:2007dq,McConnachie:2012vd,Koposov:2015cua,Bechtol:2015cbp}. For the star cluster they found the constraints from requiring that
\begin{enumerate}
\item{the timescale for the half-light radius to grow from $r_{{\rm h,} 0} = 2 \, {\rm pc}$ to the observed value, $r_{\rm h} = 13 \, {\rm pc}$, is larger than the cluster age,} 
\item{the timescale for the cluster to double in area is less than the cluster age.}
\end{enumerate}
In each case they considered two ages for the cluster, $3$ and $12 \, {\rm Gyr}$ which are at the lower and upper ends of the plausible range of values, and two different values for the dark matter density and velocity dispersion ($1$ \& $0.02 \, M_{\odot} \, {\rm pc}^{-3}$ and $5$  \& $10 \, {\rm km \, s}^{-1}$ respectively). For the ultra-faint dwarfs 
they found the constraints from requiring that 
\begin{enumerate}
\item{the timescale to grow from $r_{{\rm h,} 0} = 2 \, {\rm pc}$ to the observed value, $r_{\rm h} = 30 \, {\rm pc}$,
is less than $10 \, {\rm Gyr}$,}
\item{the timescale to double in area is less than $10 \, {\rm Gyr}$,} 
\end{enumerate}
for $\rho= 1 \, M_{\odot} \, {\rm pc}^{-3}$ and $\sigma = 5$ \& $10 \, {\rm km \, s}^{-1}$.
Apart from for the low density cases, $\rho=0.02 \, M_{\odot} \, {\rm pc}^{-3}$, the Eri II star cluster disruption constraints are tighter than the ultra-faint dwarf constraints. However, as discussed in Ref.~\cite{Brandt:2016aco}, it is possible that the star cluster constraints can be evaded, for instance if the star cluster has only recently inspiraled to the centre of Eri II, or if the star cluster's apparent position at the centre is a projection effect. The ultra-faint dwarf constraints are therefore more robust. We consider the tightest constraint from the Eri II star cluster (which comes from considering the timescale for the half-light radius to grow from $r_{{\rm h,} 0} = 2 \, {\rm pc}$ to $r_{\rm h} = 13 \, {\rm pc}$ and setting $\rho= 1 \, M_{\odot} \, {\rm pc}^{-3} $ and $\sigma = 5 \, {\rm km \, s}^{-1}$) and the weakest ultra-faint dwarf constraint (which comes from considering the timescale to grow from 
$r_{{\rm h,} 0} = 2 \, {\rm pc}$ to $r_{\rm h} = 30 \, {\rm pc}$ and setting $\rho= 1 \, M_{\odot} \, {\rm pc}^{-3} $ and $\sigma = 5 \, {\rm km \, s}^{-1}$). These constraints are shown in Fig.~\ref{fig:EROS}.

Finally, generalising the dynamical heating calculation to a non-delta function mass function, Eq.~(\ref{rh}) becomes
\begin{equation}
\label{rh2}
 \frac{{\rm d} r_{\rm h}}{{\rm d} t} = \frac{ 4 \sqrt{2} \pi G }{\sigma} \left( \alpha \frac{M_{\star}}{\rho r_{\rm h}^2} + 2 \beta r_{\rm h} \right)^{-1} 
  \int_{0}^{\infty} \psi(M) M  \ln{\Lambda} \, {\rm d} M    \,. 
\end{equation}

\section{Extended mass functions}
\label{sec:emf}

As can be seen from Fig.~\ref{fig:EROS}, together the microlensing and dynamical constraints exclude MACHOs with a delta function mass function in the mass range $10^{-3} < M/M_{\odot} <10^{4}$ making up all of the dark matter ($f=1$). However it has recently been pointed out that for mass ranges where a delta function mass function with $f=1$ is excluded an extended mass function might still satisfy all of the constraints~\cite{Carr:2016drx}. For PBHs produced from the collapse of large inflationary density perturbations an extended mass function is expected, due to the spread in masses produced by critical collapse~\cite{Niemeyer:1997mt} and also, potentially, from a spread in formation times~\cite{Kuhnel:2015vtw}. However, as emphasised in Ref.~\cite{Carr:2016drx}, an extended DHF can not be confronted with the observations by simply comparing it directly with the constraints on the halo fraction calculated for a delta function. For instance if a constraint is independent of $M$, $f(M)<f_{\rm lim}$, in the range $M_{1}$ to $M_{2}$, the integral of the DHF over this mass range must be less than $f_{\rm lim}$, and it is not possible to have $f(M) \sim f_{\rm lim}$ for MACHOs with multiple masses within this range.

Ref.~\cite{Carr:2016drx} presents a technique for applying mass dependent PBH abundance constraints calculated assuming a delta function halo fraction to extended DHFs. They divide the relevant mass range into bins. First they integrate the extended DHF within the lowest mass bin
and compare the result with the {\em weakest} bound on $f$ for a delta function halo fraction in this mass range. They then repeat the process for the first and second bins, first to third bins and so on until the whole mass range for which $f<1$ has been covered. For the constraint on the upper part of the mass range the process is reversed i.e. first they integrate the extended DHF within the highest mass bin and compare the result with the {\em weakest} bound on $f$ for a delta function mass function in this mass range. 

This process underestimates the strength of the constraint, since apart from at one edge of the bin the observational constraint is actually stronger than the considered value. If a DHF is excluded by this process it is definitely excluded, however some DHFs that are allowed by this method will in fact violate the constraint, and hence be excluded. For instance, some DHFs which are allowed when confronted with the microlensing delta function constraints using this method in fact produce $N_{\rm exp} > 3$ microlensing events, and are hence excluded. Conversely comparing the integral of the extended DHF with the tightest bound on $f$ for a delta function halo fraction would overestimate the strength of the constraint. In that case a DHF which was allowed would definitely be allowed, however DHFs that were ruled out might actually be consistent with the constraint.

To explicitly demonstrate this, we consider the EROS-2 microlensing limit in the mass range $10^{-2} < M/M_{\odot} < 10^{-1}$, where the constraint on the halo fraction for a delta function mass function is tightest and has the weakest mass dependence.  The requirement that there are $N_{\rm exp} < 3$ microlensing events in the EROS-2 survey leads to the limit $f<0.036$ for $M=10^{-2} M_{\odot}$ , while for 
$M=10^{-1} M_{\odot}$, $f<0.053$. We consider a top-hat extended mass function which is flat for $10^{-2} < M/M_{\odot} < 10^{-1}$ and zero elsewhere. If the amplitude is fixed so that $f=0.053$ i.e.~to match the weakest constraint on $f$ in this mass range (as in the method presented in Ref.~\cite{Carr:2016drx}) then the number of microlensing events expected is $N_{\rm exp} =3.6$, which exceeds the limit. This confirms that that the method presented in Ref.~\cite{Carr:2016drx} under-estimates the strength of the constraint.  The largest halo fraction allowed for the top-hat extended DHF is in fact $f=0.044$, which is close to the mean of the values for a delta function mass function with mass at the upper and lower ends of the mass range considered. 

Using the weakest value of the delta function limit leads to larger errors in mass regions where the delta function halo fraction limit varies rapidly with mass. 
It might be possible to devise a reliable way of applying the limits calculated for a delta function to extended DHFs in these regions, for instance by comparing with the appropriately mass-weighted average of the delta function limit. However to definitively ascertain whether or not a DHF is consistent with a given constraint, it is necessary to explicitly recalculate the constraint for that DHF. 

\begin{figure}
\includegraphics[width=0.45\textwidth]{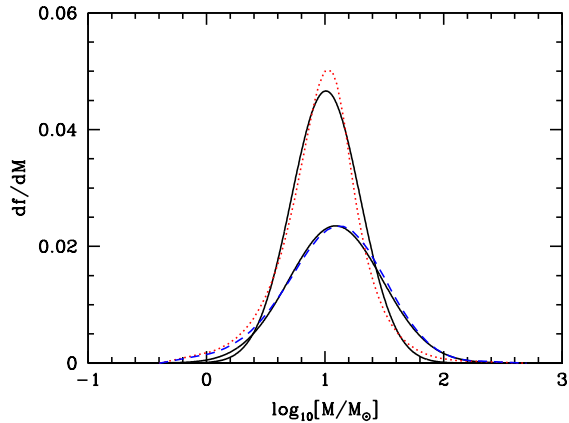}
\caption{\label{fig:mf} The PBH DHF, ${\rm d} f/{\rm d} M$, for the axion-curvaton (red dotted line) and running mass (blue dashed) inflation models from Ref.~\cite{Carr:2016drx}. The black lines are the least squares fit of the functional form Eq.~(\ref{lnmf}) to these DHFs. In all cases the DHFs integrate to unity (i.e. all of the halo dark matter is in the form of PBHs).}
\end{figure}

Fig.~6 of Ref.~\cite{Carr:2016drx} shows two PBH DHFs, originating from axion curvaton (AC) and running-mass (RM) inflation, which they find satisfy the EROS-2 microlensing and dynamical constraints using their method. Using the expressions in Sec.~\ref{sec:micro}, for the number of microlensing events produced by an extended DHF, we find that these DHFs would have produced 5.5 (AC) and 4.1 (RM) microlensing events in the EROS-2 survey.  Therefore they are both excluded by the EROS-2 microlensing constraint alone. Integrating Eq.~(\ref{rh2}) we find that these extended DHFs would produce excessive heating of the Eri II star cluster within $1.1$ (AC) and  $0.7 \, {\rm Gyr}$ (RM) and of ultra-faint dwarf galaxies within $5.8$ (AC) and  $3.8. \, {\rm Gyr}$ (RM). Therefore these two DHFs each produce too many microlensing events and also excessively heat ultra-faint dwarfs, and are hence excluded by both constraints individually.

\begin{figure}
\includegraphics[width=0.45\textwidth]{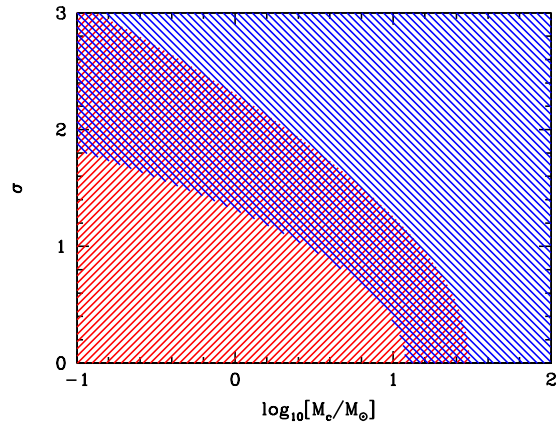}
\caption{\label{gaussnexp3} Constraints on the width, $\sigma$, of the DHF functional form, eq.~(\ref{lnmf}), as a function of the central mass $M_{\rm c}$. Parameter values in the red hatched area in the bottom left produce $N_{\rm exp}  \geq 3$ microlensing events in the EROS-2 survey and are excluded at $95\%$ confidence. The blue hatched area in the top right is excluded by the heating of ultra-faint dwarfs. The constraint from the disruption of the star cluster in Eri II is tighter and excludes a large region of parameter space}
\end{figure}

In order to explore whether extended DHFs can satisfy both the microlensing and dynamical constraints we consider a functional form for the DHF
\begin{equation}
\label{lnmf}
\psi(M) \equiv \frac{{\rm d} f}{{\rm d} M} =  N \exp{ \left[ -  \frac{ (\log{M} - \log{M_{\rm c}})^2}{2 \sigma^2} \right]} \,.
\end{equation}
where $N$ is a normalisation constant chosen so that the DHF is normalised to unity~\footnote{A log-normal distribution only differs from this functional form at the per-cent level for the values of $M$ for which ${\rm d} f/{\rm d} M$ is non-negligible, and does not provide a better fit to the DHFs from Ref.~\cite{Carr:2016drx}. It also has the disadvantage that the value of its mode depends on both $M_{\rm c}$ and $\sigma$.}.
The least squares fits of Eq.~(\ref{lnmf}) to the axion curvaton and running-mass inflation DHFs from Ref.~\cite{Carr:2016drx} are shown in Fig.~\ref{fig:mf}. The numbers of microlensing events and disruption time-scales for the best-fit functional forms differ from those of the original DHFs by less than $10\%$, indicating that Eq.~(\ref{lnmf}) is a reasonable approximation to the PBH DHF produced by these inflation models.

\begin{figure}
\includegraphics[width=0.45\textwidth]{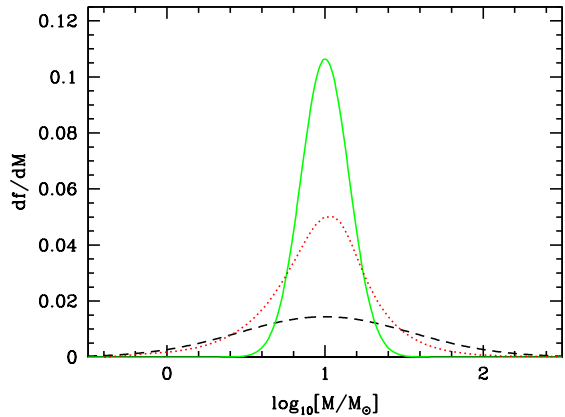}
\caption{\label{microdisrupt} The differential PBH halo fraction, ${\rm d} f/{\rm d} M$, for the axion-curvaton (dotted red line) inflation model from Ref.~\cite{Carr:2016drx} compared with the broadest differential halo fraction, centered at the same mass, which satisfies the ultra-faint dwarf disruption constraint (solid green) and the narrowest differential halo fraction which satisfies the EROS-2 microlensing constraint (dashed black).}
\end{figure}

 Fig.~\ref{gaussnexp3} shows the $\sigma$ and $M_{\rm c}$ values excluded by microlensing and the weakest ultra-faint dwarf heating constraint.
The requirement that the DHF produces $N_{\rm exp}  \leq 3$ microlensing events in the EROS-2 survey places a lower limit on $\sigma$ for $M_{\rm c} < 30 \, M_{\odot}$, i.e. narrow DHFs are excluded. On the other-hand the dynamical heating constraints place an upper limit on $\sigma$ (i.e. broad halo DHFs are excluded), and for the weakest ultra-faint dwarf constraint a delta function mass function is excluded for $M> 12 M_{\odot}$. These two competing constraints on the width of the DHF overlap, and there is no width and central mass for which both constraints are satisfied. For the tightest Eri II star cluster constrain a delta function halo function is excluded for $M> 1.5 M_{\odot}$ and the overlap between the microlensing and dynamical constraints is even larger.

To illustrate the conflict between the constraints, in Fig.~\ref{microdisrupt} we compare the PBH DHF from the axion curvaton inflation model from Ref.~\cite{Carr:2016drx}, with the broadest DHF (centered at the same mass) which satisfies the ultra-faint dwarf disruption constraint and the narrowest DHF which satisfies the EROS-2 microlensing constraint. The axion curvaton DHF is significantly broader than the broadest DHF which satisfies the ultra-faint dwarf disruption constraint and significantly narrower than the narrowest DHF which satisfies the EROS-2 microlensing constraint. It is therefore clearly excluded by both constraints.

\section{Discussion}
\label{sec:discussion}

Microlensing surveys~\cite{Tisserand:2006zx,Alcock:1998fx} constrain the halo fraction of MACHOs with $10^{-7} <M /M_{\odot} < 10$, while dynamical heating constraints are sensitive to $M/M_{\odot} \gtrsim 10$~\cite{Brandt:2016aco}. Together they exclude MACHOs with $10^{-7} <M /M_{\odot} < 10^{5}$ and a delta function mass function from making up all of the dark matter. However Refs.~\cite{Clesse:2015wea,Carr:2016drx} have pointed out that MACHOs with an extended mass function, as expected for PBHs formed from the collapse of large inflationary density perturbations~\cite{Niemeyer:1997mt,Kuhnel:2015vtw}, might be compatible with these constraints. Furthermore interest in PBHs with $M \sim 10 M_{\odot}$~\cite{Bird:2016dcv,Clesse:2016vqa,Sasaki:2016jop} has recently been stimulated by the discovery of gravitational waves from massive black hole mergers by LIGO~\cite{Abbott:2016blz}.

We have explicitly calculated the microlensing and dynamical constraints for the DHFs found for PBHs produced by two inflation models in Ref.~\cite{Carr:2016drx} and also for a variable width DHF which replicates their shape. The DHFs studied in 
Ref.~\cite{Carr:2016drx} both produce $N_{\rm exp} > 3$ microlensing events in EROS-2 and also cause excessive dynamical heating of ultra-faint dwarf galaxies. In general we find that the dynamical constraints place a (central mass dependent) upper limit on the width of the DHF (i.e.~wide distributions are excluded), while the microlensing constraints place a (central mass dependent) lower limit on the width of the DHF (i.e.~narrow distributions are excluded). These constraints overlap and there are no parameter values which satisfy both the microlensing limt and the weakest ultra-faint dwarf heating limit.

We have not proved that there is no extended DHF, with all of the dark matter in compact objects in a single mass range,  which can satisfy both the microlensing and dynamical constraints. However we have shown that
\begin{itemize}
\item to ascertain whether an extended DHF satisfies the microlensing and dynamical constraints it is necessary to recalculate the limits for the specific mass function, rather than using the limits derived for a delta function mass function,
\item generic DHFs, which replicate the PBH distributions produced by inflation models, can not simultaneously satisfying the EROS-2 microlensing constraint and also the weakest ultra-faint dwarf heating limit.  
\end{itemize}

\vspace*{1cm}
\acknowledgments

A.M.G.  acknowledges  support  from  STFC  grant ST/L000393/1 and is grateful to Marit Sandstad for useful comments that have improved the presentation of the manuscript.

\end{document}